\documentclass[journal]{IEEEtran}
\usepackage{setspace}

\usepackage{epsfig}
\usepackage{amsmath}
\usepackage{amsfonts}
\usepackage{color}

\vspace{-20mm}
\begin{document}
\vspace{-20mm}
\title{Optimal Power Allocation and Secrecy Sum Rate \\in Two-Way Untrusted Relaying}
\author{Ali Kuhestani$^1$, Phee Lep Yeoh$^2$, and Abbas Mohammadi$^1$\\
 $^1$Electrical Engineering Department, Amirkabir University of Technology, Tehran, Iran.\\
 $^2$School of Electrical and Information Engineering, The University of Sydney, NSW, Australia.}

\markboth{}{Ali Abbas}
\maketitle
{\vspace{-10mm}}
\begin{abstract}
In this paper, we examine the secrecy performance of two-way relaying between a multiple antenna base station (BS) and a single antenna mobile user (MU) in the presence of a multiple antenna friendly jammer (FJ).  We consider the untrusted relaying scenario where an amplify-and-forward relay is both a necessary helper and a potential eavesdropper. To maximize the instantaneous secrecy sum rate, we derive new closed-form solutions for the optimal power allocation (OPA) between the BS and MU under the scenario of relaying with friendly jamming (WFJ). Based on the OPA solution, new closed-form expressions are derived for the ergodic secrecy sum rate (ESSR) with Rayleigh fading channel. Furthermore, we explicitly determine the high signal-to-noise ratio slope and power offset of the ESSR to highlight the benefits of friendly jamming. Numerical examples are provided to demonstrate the impact of the FJ's location and number of antennas on the secrecy performance.
\end{abstract}
{\vspace{-2mm}}
\section{Introduction}
\IEEEPARstart{R}{elaying} is a proven approach to improve energy efficiency, extend coverage and increase the throughput of wireless communication networks. Recently, the benefits of relaying have been viewed from the viewpoint of wireless physical-layer security (PLS)~\cite{yener15}. A key area of interest is the untrusted relaying scenario where the source-to-destination transmission is assisted by a relay which may also be a potential eavesdropper~{\cite {review3}}. This scenario occurs in large-scale wireless systems such as heterogeneous networks and device-to-device (D2D) communications, where confidential messages are often retransmitted by intermediate nodes.

Secure transmission utilizing an untrusted relay was first studied in {\cite {he11}}, where an achievable secrecy rate was derived. In {\cite {he1}}, it was found that introducing a friendly jammer (FJ) could result in a positive secrecy rate for a one-way untrusted relay link with no direct source-destination transmission. Indeed, many recent papers on untrusted relay communications have focussed on the one-way relaying scenario {\cite {sun}}-- {\cite {Kuhestani2}}. Recently, several works have considered the more interesting scenario of two-way untrusted relaying {\cite {zhang0}}-{\cite {Tao2}} where physical-layer network coding can provide security enhancement since the relay receives a superimposed signal from the two sources instead of each individual signal {\cite{Koosha}}.

Considering two-way untrusted relaying, {\cite {zhang0}} proposed a game-theoretic power control scheme between the two single antenna sources and multiple FJs. In the proposed scheme, the optimal jamming power was derived for the special case where the FJ is very close to the amplify-and-forward (AF) relay. In {\cite {Huang3}}, multiple antennas were considered at the sources and a joint optimization of transmit covariance matrices and relay selection was proposed for AF relaying without FJs. In {\cite {Tao2}}, an iterative algorithm was proposed to jointly optimize the multiple antenna sources and AF relay beamformers such that the instantaneous secrecy sum rate without FJs is maximized.

In this paper, we investigate the PLS of a two-way untrusted AF relaying system, where a multiple antenna base station (BS) exchanges confidential messages with a single antenna mobile user (MU) in the presence of a multiple antenna FJ. The relay is considered to be both a necessary helper and a potential curious eavesdropper. For this system, we analyze the secrecy performance under the scenario of relaying with friendly jamming (WFJ). In particular, we formulate the optimal power allocation (OPA) between the BS and MU that maximizes the instantaneous secrecy sum rate of two-way untrusted relaying. Based on this, we derive a new closed-form solution for the OPA in the large-scale multiple antenna (LSMA) regime for the BS and FJ. According to our OPA solution, the ergodic secrecy sum rate (ESSR) of the optimized system is derived for Rayleigh fading channel. We further characterize the high signal-to-noise ratio (SNR) slope and the high SNR power offset of the ESSR which explicitly captures the impact of the distance dependent channel gains and the number of BS and FJ antennas on the ESSR. Finally, numerical results are provided to reveal the secrecy performance advantage of the OPA scheme compared with equal power allocation and without friendly jamming. We highlight the significant ESSR advantage of employing a FJ and the impacts of the FJ location and number of BS and FJ antennas on the ESSR.

{\vspace{-2mm}}
\section{ System Model} \label{Sec Model}
We consider a two-way relay network as depicted in Fig. 1 where a BS and MU exchange information via an untrusted AF relay. Depending on the required secrecy sum rate for the BS and MU, a FJ may be employed to transmit noise-like jamming signals to the relay. The BS and FJ are equipped with $N_\mathrm{BS}$ and $N_\mathrm{FJ}$ antennas, respectively, while the relay and MU are equipped with a single antenna. Similar to {\cite {zhang0}}, we consider that the FJ is an external power seller that may be activated to boost the secrecy sum rate of the BS and MU.

The two-way relaying transmission is performed in two phases. In Phase 1, shown with solid lines, both the BS and MU transmit their information to the relay. Simultaneously, the FJ can be activated to transmit jamming signals to the relay. In Phase 2, shown with dashed lines, the relay broadcasts a combined version of the received signals to the BS and MU. Based on knowledge of its own signal and the FJ's jamming signal, the BS and MU extracts the information signal from its opposite counterpart. To utilize the multiple antennas, we apply maximum ratio transmission (MRT) at the BS and FJ in Phase 1, and maximum ratio combining (MRC) at the BS in Phase 2.

\begin{figure}[t] \label{1}
  \begin{center}
    \includegraphics[width=3.3in,height=2.8in]{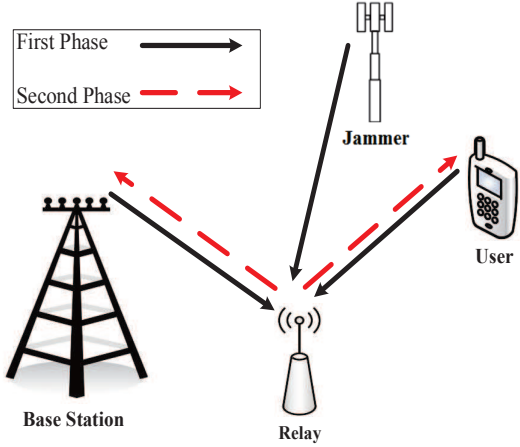} 
    \vspace{-8mm}\caption{Two-way relaying between a multiple antenna base station (BS) and a mobile user (MU) via an untrusted amplify-and-forward (AF) relay in the presence of a multiple antenna friendly jammer (FJ).}
  \label{fig1}\end{center}
  {\vspace{-5mm}}
\end{figure}

We consider a total transmit power budget for the BS and MU of $P$ with power allocation factor $\lambda\in (0,1)$ such that the transmit powers at the BS and MU are $\lambda P$ and $(1-\lambda)P$, respectively. For simplicity, the transmit power at the relay and the FJ are set to $P$. The complex Gaussian channel vectors from the BS to relay, MU to relay and {FJ to relay} are denoted by ${\bf h}_{\mathrm{br}}\sim \mathcal{CN}({\bf 0}_{{N_{\mathrm{BS}}}\times 1},\mu_{{\mathrm{br}}}{\bf {I}}_{N_{\mathrm{BS}}\times 1})$, ${h}_{\mathrm{mr}}\sim \mathcal{CN}( 0 ,\mu_{{\mathrm{mr}}})$ and ${\bf h}_{\mathrm{fr}}\sim \mathcal{CN}({\bf 0}_{{N_{\mathrm{FJ}}}\times 1},\mu_{{\mathrm{fr}}}{\bf {I}}_{N_{\mathrm{FJ}}\times 1})$, respectively, where  ${\bf 0}$ is the zero matrix, $\bf {I}$ is the identity matrix, and $\mu_{\mathrm{br}}$, $\mu_{\mathrm{mr}}$, $\mu_{\mathrm{fr}}$ are the channel gains based on the distance-dependent path loss for each antenna branch. We define $\gamma_{\mathrm{br}}=\rho\|{\bf h}_{\mathrm{br}}\|^2$, $\gamma_{\mathrm{mr}}=\rho|{h}_{\mathrm{mr}}|^2$ and $\gamma_{\mathrm{fr}}={\rho}\|{\bf h}_{\mathrm{fr}}\|^2$  where $\rho=\frac{P}{\sigma_n^2}$ and the additive white Gaussian noise (AWGN) at all nodes are zero-mean with variance ${\sigma_n^2}$. Also $\overline{\gamma}_{\mathrm{br}}=\rho \mu_{\mathrm{br}}$, $\overline{\gamma}_{\mathrm{mr}}=\rho \mu_{\mathrm{mr}}$ and $\overline{\gamma}_{\mathrm{fr}}={\rho} \mu_{\mathrm{fr}}$.

Let ${x}_1$ and ${x_2}$ denote the transmitted scalar symbols from the BS and MU, respectively. In Phase 1, the received signal at the relay is expressed as
\begin{eqnarray}\label{yy_r}
y_{\mathrm{R}}{\hspace{-.7mm}}={\hspace{-.7mm}}\sqrt{\lambda P} {{\Arrowvert {\bf h}_{{\mathrm{br}}}\Arrowvert}}{x}_1{\hspace{-1mm}}+{\hspace{-1mm}}\sqrt{(1-\lambda)P}  h_\mathrm{mr} { x}_{2}{\hspace{-1mm}}+{\hspace{-1mm}}\sqrt{ {P}} {{\Arrowvert {\bf h}_{{\mathrm{fr}}}\Arrowvert}}x_\mathrm{fj}{\hspace{-.5mm}}+{\hspace{-.5mm}}n_\mathrm{R},\hspace{-2mm}
\end{eqnarray}
where $x_\mathrm{fj}$ and $n_{\mathrm{r}}$ represent the jamming signal and the AWGN at the relay, respectively.

In Phase 2, the relay amplifies its received signal in \eqref{yy_r} by an amplification factor of
\begin{equation}\label{relaygain}
G=\sqrt{\frac{P}{\lambda P {{\Arrowvert {\bf h}_{{\mathrm{br}}}\Arrowvert}}^2+(1-\lambda)P|h_\mathrm{mr}|^2+ {P} {{\Arrowvert {\bf h}_{{\mathrm{fr}}}\Arrowvert}}^2+\sigma_n^2}},
\end{equation}
and transmits a broadcast signal to the BS and MU given by $x_{\mathrm{R}}=Gy_{\mathrm{R}}$. After canceling the self interference and jamming signals in $y_{\mathrm{R}}$, the corresponding received signals at the BS and MU are given by
\begin{eqnarray}\label{yy_d}
{\bf y}_\mathrm{BS}{\hspace{-3mm}}&=&{\hspace{-3mm}}\sqrt{(1-\lambda)P}G{\bf h}_\mathrm{br} h_\mathrm{mr} x_2+G {\bf h}_{\mathrm{br}}n_\mathrm{R} + n_\mathrm{BS},\\\label{yy_d2}
{y}_\mathrm{MU}{\hspace{-3mm}}&=&{\hspace{-3mm}}\sqrt{\lambda P}G {{\Arrowvert {\bf h}_{{\mathrm{br}}}\Arrowvert}}h_\mathrm{mr}{x}_1+Gh_{\mathrm{mr}}n_\mathrm{R} + n_\mathrm{MU},
\end{eqnarray}
respectively, where $n_\mathrm{BS}$ and $n_{\mathrm{MU}}$ are the AWGN at the BS and MU, respectively.

Based on (\ref{yy_d}) and \eqref{yy_d2}, and using (\ref {relaygain}), the signal-to-interference-plus-noise ratios (SINRs) at the BS (after performing MRC) and at the MU are given by
\begin{eqnarray}\label{gammas}
\gamma_{\mathrm{BS}}{\hspace{-3mm}}&=&{\hspace{-3mm}}\frac{(1-\lambda)\gamma_{\mathrm{br}}\gamma_{\mathrm{mr}}}{(1+\lambda)
\gamma_{\mathrm{br}}+(1-\lambda)\gamma_{\mathrm{mr}}+ \gamma_\mathrm{fr}+\epsilon},\\ \label{gammas2}
\gamma_{\mathrm{MU}}{\hspace{-3mm}}&=&{\hspace{-3mm}}\frac{\lambda\gamma_{\mathrm{br}}
\gamma_{\mathrm{mr}}}{\lambda\gamma_{\mathrm{br}}+(2-\lambda)\gamma_{\mathrm{mr}}+\gamma_\mathrm{fr}+\epsilon},
\end{eqnarray}
respectively, where $\epsilon=1$. Based on (\ref{yy_r}), the SINR at the untrusted relay is given by
\begin{eqnarray}\label{gammaRelay}
\gamma_{\mathrm{R}}\hspace{-3mm}&=&\hspace{-3mm}\frac{\lambda \gamma_{\mathrm{br}} + (1-\lambda) \gamma_{\mathrm{mr}}}{\gamma_{\mathrm{fr}}+\epsilon},
\end{eqnarray}
where we assume that the relay performs multiuser decoding to estimate the signals from the BS and MU. We adopt the well-known SINR approximation of $\epsilon=0$~ and note that from a security perspective this corresponds to the maximum intercept probability for the relay SINR in (\ref {gammaRelay}). From (\ref {gammas})--(\ref {gammaRelay}), we can conclude that the jamming signal decreases all the SINR terms but as observed in the next section, it has a more dominant impact on the relay SINR.

The instantaneous secrecy sum rate of the two-way untrusted relaying system is given by {\cite {Huang3}} $R_s=[I_{\mathrm{BS}}+I_{\mathrm{MU}}-I_\mathrm{R}]^+$, where  $I_{{K}}=\frac{1}{2}\log_2(1+\gamma_{{K}})$ is the achievable rate at source $K \in \{\mathrm{BS}~,\mathrm{MU}\}$, $I_{\mathrm{R}}=\frac{1}{2}\log_2(1+\gamma_{\mathrm{R}})$ is the information leaked to the untrusted relay, and  $[x]^+ = \max \{ 0, x \}$. As such, the instantaneous secrecy sum rate can be re-expressed as $R_s=\Big[\frac{1}{2}\log_2\Phi (\lambda)\Big]^+$, where
\begin{equation}\label{phi}
\Phi(\lambda)=\frac{(1+\gamma_{\mathrm{BS}})(1+\gamma_{\mathrm{MU}})}{1+\gamma_\mathrm{R}}.
\end{equation}
\section{Optimal Power Allocation for Secrecy Sum Rate Maximization}
{\vspace{-.5mm}}
In this section, we derive the OPA at the BS and MU that maximizes the instantaneous secrecy sum rate. Indeed, we need to consider the following optimization problem
\begin{eqnarray}\label{opt_prob_fj}
\lambda^{\star}=\mathrm{arg}~\max  \Phi (\lambda), ~~{\mathrm{s.t.}}~0<\lambda < 1
\end{eqnarray}
Note that $\Phi(\lambda)$ is a quasi-concave function of $\lambda$ in the feasible set. As such, by taking the
derivative of $\phi(\lambda)$ w.r.t. $\lambda$ and setting it to zero, the OPA when ${\gamma_{\mathrm{br}}} \gg {\gamma_{\mathrm{mr}}}$ is derived as
\begin{eqnarray}\label{sol_J0}
\lambda^{\star}{\hspace{-1mm}}={\hspace{-1mm}}\frac{-2\gamma_\mathrm{mr}{\hspace{-1mm}}-{\hspace{-1mm}}\gamma_\mathrm{fr}{\hspace{-1mm}}+{\hspace{-1mm}}
\gamma_\mathrm{mr}\sqrt
{2\gamma_\mathrm{mr}^2{\hspace{-1mm}}+{\hspace{-1mm}}3\gamma_{\mathrm{mr}}\gamma_\mathrm{fr}{\hspace{-1mm}}-{\hspace{-1mm}}
2\gamma_\mathrm{mr}{\hspace{-1mm}}+{\hspace{-1mm}}\gamma_\mathrm{fr}^2{\hspace{-1mm}}-{\hspace{-1mm}}\gamma_\mathrm{fr}}}
{\gamma_\mathrm{br} \gamma_\mathrm{mr}}.
\end{eqnarray}
To obtain further insights from \eqref{sol_J0}, we consider the case where $\gamma_\mathrm{fr}\gg \gamma_\mathrm{mr}$ and $\gamma_\mathrm{mr} \gg 1$ which results in
\begin{eqnarray}\label{sol_J2}
\lambda^{\star}=\frac{\gamma_\mathrm{fr}}{\gamma_\mathrm{br}}.
\end{eqnarray}
This OPA applies to the case of high SNRs in both the relay-to-jammer link and the relay-to-MU link. Note that $\gamma_\mathrm{fr}\gg \gamma_\mathrm{mr}$ occurs when the number of antennas at the FJ $N_\mathrm{FJ}$ is significantly large, or the relay is located much closer to the FJ compared to the MU. It is also worth of noting that while we have considered LSMA at both the BS and the FJ in deriving \eqref{sol_J2}, $N_\mathrm{BS}$ should be larger than $N_\mathrm{FJ}$ to ensure that $\lambda<1$.
{\vspace{-.2mm}}
\section{Ergodic Secrecy Sum Rate Analysis}
{\vspace{-.2mm}}
In this section, we apply the OPA solution (\ref {sol_J2}) to derive new closed-form expressions for the ESSR of two-way untrusted relaying. Recall that the ESSR can be expressed as {\cite {Huang3}}
\begin{eqnarray}
\overline{R_s}{\hspace{-3mm}}&=&{\hspace{-3mm}}\frac{1}{2 \ln 2}\Big[\underbrace{\mathbb{E}\Big\{\ln (1+\gamma_\mathrm{BS})\Big\}}_{I_1}+\underbrace{\mathbb{E}\Big\{\ln (1+\gamma_\mathrm{MU})\Big\}}_{I_2}\nonumber\\
&-&{\hspace{-3mm}}\underbrace{\mathbb{E}\Big\{\ln (1+\gamma_\mathrm{R})\Big\}}_{I_3}\Big].
\end{eqnarray}
As such, in the following subsections, we proceed to evaluate the ESSR terms of $I_1$, $I_2$ and $I_3$.

Substituting $\lambda^{\star}$ in (\ref {sol_J2}) into (\ref {gammas})--(\ref {gammaRelay}), the SINRs at the BS, MU and relay for ${\gamma_{\mathrm{br}}} \gg {\gamma_{\mathrm{mr}}}$  and $\gamma_\mathrm{mr} \gg 1$ yields
\begin{align}\label{SNR_Large}
\gamma_{\mathrm{BS}}=\gamma_\mathrm{mr}\frac{1-\frac{\gamma_\mathrm{fr}}{\gamma_\mathrm{br}}}
{1+ \frac{2\gamma_\mathrm{fr}}{\gamma_\mathrm{br}}},~~~\gamma_\mathrm{MU}= \frac{\gamma_\mathrm{mr}}{2}~~~\mathrm{and}~~~\gamma_\mathrm{R}=1.
\end{align}

To derive the ESSR term of $I_1$, we use {\it Lemma 3} in {\cite {Alkheir}} where the exponential integral function $\mathrm{Ei}(-x)$ {\cite {table}} is expressed as a closed-form solution as
\begin{align}\label{Ei_x}
\mathrm{Ei}(-x)\approx -4 \sqrt{2} \pi a_1 a_2 \sum_{p=1}^{T+1}\sum_{q=1}^{T'+1}\sqrt{b_p}e^{-4b_pb_qx},
\end{align}
where $a_1=\frac{1}{2(T+1)}$, $a_2=\frac{1}{2(T'+1)}$, and $b_{p(q)}=\frac{\mathrm{cot}(\theta_{p-1(q-1)})-\mathrm{cot}(\theta_{p(q)})}{\theta_{p(q)}-\theta_{p-1(q-1)}}$, and  $.065<\theta_1<...<\theta_{T+1(T'+1)}$. $T$ and $T'$ are positive integers that control the accuracy of the approximation.

Using (\ref {SNR_Large}) and (\ref {Ei_x}), $I_1$ is obtained as (see Appendix \ref{appC}){\vspace{-1mm}}
\begin{eqnarray}\label{ESSR_large}
I_1{\hspace{-2.7mm}}&=&{\hspace{-2.7mm}}\mathbb{E}\Big\{\ln(1+\gamma_\mathrm{BS})\Big\}\nonumber\\
&=&{\hspace{-2.7mm}}4\sqrt{2} \pi a_1 a_2~ e^{\frac{(4 b_p b_q -1)({2-\frac{3}{\varrho}})}{{\overline{\gamma}_\mathrm{mr}}}}~\frac{\Gamma(N_\mathrm{FJ}+N_\mathrm{BS})}{\Gamma(N_\mathrm{FJ})\Gamma(N_\mathrm{BS})} \Big(\frac{\overline{\gamma}_\mathrm{fr}}{\overline{\gamma}_\mathrm{br}}\Big)^{N_\mathrm{BS}}\nonumber\\
&\times&{\hspace{-3.5mm}} \sum_{p=1}^{T+1}\sum_{q=1}^{T'+1}\sqrt {b_p} \Big[-A_1 ~ \mathrm{Ei}\Big(\frac{(4 b_p b_q -1)({3-\frac{3}{\varrho}})}{{\overline{\gamma}_\mathrm{mr}}}\Big)\nonumber\\
&+&{\hspace{-3.5mm}}\sum_{i=1}^{N_\mathrm{FJ}+N_\mathrm{BS}-1}{\hspace{-2.7mm}}A_{i+1}\Big(-\frac{(1-4b_pq_p)^i \mathrm{Ei}(-\frac{(4 b_p b_q -1)({3-\frac{3}{\varrho}})}{{\overline{\gamma}_\mathrm{mr}}})}{i!}\nonumber\\
&+&{\hspace{-2.7mm}} \frac{e^{-\frac{(4 b_p b_q -1)({3-\frac{3}{\varrho}})}{{\overline{\gamma}_\mathrm{mr}}}}}{(\frac{{3-\frac{3}{\varrho}}}{{\overline{\gamma}_\mathrm{mr}}})^i}\sum_{k=0}^{i-1}
\frac{(1-4b_pb_q)^k(3-\frac{3}{\varrho})^k}{i(i-1)...(i-k)}\Big)\Big],{\vspace{-2mm}}
\end{eqnarray}
where $\varrho=1+\frac{\overline{\gamma}_\mathrm{fr}}{\overline{\gamma}_\mathrm{br}}$ and  $A_{N_\mathrm{FJ}+N_\mathrm{BS}+1-i}=\frac{1}{(i-1)!}\frac{\partial^{i-1} \Big[u^{N_\mathrm{BS}-1}(u-1)^{N_\mathrm{FJ}-1}\Big]}{\partial u^{i-1}}|_{u=\frac{1}{\varrho}}$. According to $\gamma_\mathrm{MU}$ in (\ref {SNR_Large}) and using [11, Eq. (4.331.2)], $I_2$ can be obtained as
\begin{eqnarray}\label{ESSR_large2}
I_2=\mathbb{E}\Big\{\ln(1+\gamma_\mathrm{MU})\Big\}=-e^{-\frac{2}{\overline{\gamma}_\mathrm{mr}}}
\mathrm{Ei}\Big(-\frac{2}{\overline{\gamma}_\mathrm{mr}}\Big),
\end{eqnarray}
and finally, using $\gamma_\mathrm{R}$ in (\ref {SNR_Large}), $I_3$ is given by
\begin{eqnarray}\label{ESSR_large3}
I_3=\mathbb{E}\Big\{\ln(1+\gamma_\mathrm{R})\Big\}=\ln 2.
\end{eqnarray}

We can further evaluate the asymptotic ESSR when $\rho\rightarrow\infty$ by applying the general asymptotic form given by {\cite {offset}}
\begin{eqnarray}\label{asymp}
\overline{R_s^\infty}=S_{\infty}\Big(\log_2 \rho-L_{\infty}\Big),
\end{eqnarray}
where $S_{\infty}$ is the high SNR slope in bits/s/Hz/ (3 dB) and $L_{\infty}$ is the high SNR power offset in 3 dB units, which are defined respectively as
\begin{eqnarray}\label{power_slope}
S_{\infty}=\lim_{\rho \rightarrow \infty}\frac{\overline{R_s^{\infty}}}{\log_2 \rho}~~~\mathrm{and}~~~L_{\infty}=\lim_{\rho \rightarrow \infty}\Big(\log_2 \rho-\frac{\overline{R_s^{\infty}}}{S_{\infty}}\Big).
\end{eqnarray}

Based on (\ref {SNR_Large}), in the high SNR regime with $\rho\rightarrow\infty$, we have $\ln (1+\gamma_\mathrm{BS})\approx \ln (\gamma_\mathrm{BS})$ and
$\ln (1+\gamma_\mathrm{MU})\approx \ln (\gamma_\mathrm{MU})$. As such, we can evaluate $I_1$ and $I_2$ as follows
\begin{eqnarray}
I_1{\hspace{-3mm}}&=&{\hspace{-3mm}}\underbrace{\mathbb{E}\Big\{\ln(\gamma_\mathrm{mr})\Big\}}_{I_{1,1}}+\
\underbrace{\mathbb{E}\Big\{\ln(1-\frac{\gamma_\mathrm{fr}}{\gamma_\mathrm{br}})\Big\}}_{I_{1,2}}-
\underbrace{\mathbb{E}\Big\{\ln(1+2\frac{\gamma_\mathrm{fr}}{\gamma_\mathrm{br}})\Big\}}_{I_{1,3}},\nonumber\\{\vspace{-4mm}}
I_2{\hspace{-3mm}}&=&{\hspace{-3mm}}\mathbb{E}\Big\{\ln(\gamma_\mathrm{MU})\Big\}=
\underbrace{\mathbb{E}\Big\{\ln({\gamma_\mathrm{mr}})\Big\}}_{I_{1,1}}-\ln 2,
\end{eqnarray}{\vspace{-1mm}}
where the values of $I_{1,1}$, $I_{1,2}$ and $I_{1,3}$ are (see Appendix \ref{appD}){\vspace{-1mm}}
\begin{eqnarray}\label{I_11}
I_{1,1}{\hspace{-2.7mm}}&=&{\hspace{-2.7mm}}\ln(\overline{\gamma}_\mathrm{mr})-\mathcal{C},\\
I_{1,2}{\hspace{-2.7mm}}&=&{\hspace{-2.7mm}}\frac{\Gamma(N_\mathrm{FJ}+N_\mathrm{BS})}{\Gamma(N_\mathrm{FJ})\Gamma(N_\mathrm{BS})}
\Big(\frac{\overline{\gamma}_\mathrm{fr}}{\overline{\gamma}_\mathrm{br}}\Big)^{N_\mathrm{BS}}\times\frac{1}{(-1)^{N_\mathrm{BS}+1}}\nonumber\\
&\times& {\hspace{-2.7mm}}\Big( B_1 \mathrm{Li}_2(\frac{\varrho-1}{\varrho})+\sum_{i=2}^{N_\mathrm{FJ}+N_\mathrm{BS}}{\hspace{-3mm}}B_i\Big[C_1 \ln(\frac{\varrho-1}{\varrho}) \nonumber\\
&+&{\hspace{-2.7mm}}\sum_{j=2}^{i-1}\frac{C_j}{1-j}\Big((1-\varrho)^{1-j}-(-\varrho)^{1-j}\Big)\Big]\Big),\label{I12}
\end{eqnarray}
and
\begin{eqnarray}\label{I13}
I_{1,3}{\hspace{-2.7mm}}&=&{\hspace{-2.7mm}}\frac{\Gamma(N_\mathrm{FJ}+N_\mathrm{BS})}{\Gamma(N_\mathrm{FJ})\Gamma(N_\mathrm{BS})}
\Big(\frac{\overline{\gamma}_\mathrm{fr}}{\overline{\gamma}_\mathrm{br}}\Big)^{N_\mathrm{BS}}\nonumber\\
&\times&{\hspace{-2.7mm}}\Big(D_1 \Big[\ln (3) \ln (\frac{2 \varrho}{2 \varrho -3})+\mathrm{Li}_2(\frac{2 \varrho}{2 \varrho -3})-\mathrm{Li}_2(\frac{2 \varrho-2}{2 \varrho -3})\Big] \nonumber\\
&+&{\hspace{-2.7mm}}\sum_{i=2}^{N_\mathrm{FJ}+N_\mathrm{BS}}\frac{2^{i-1}D_i}{i-1}\Big[-\frac{\ln (3)}{(2\varrho)^{i-1}}+\frac{\ln (3)}{(2 \varrho-3)^{i-1}}\nonumber\\
&+&{\hspace{-2.7mm}}\frac{B_j}{1-j}\Big((2\varrho)^{1-j}
-(2\varrho-2)^{1-j}\Big)\Big]\Big),
\end{eqnarray}
where $B_{N_\mathrm{FJ}+N_\mathrm{BS}+1-i}=\frac{(N_\mathrm{FJ}-1)(N_\mathrm{FJ}-2)....(N_\mathrm{FJ}+1-i)}{(i-1)!}(\varrho-1)^{N_\mathrm{FJ}-i}$, $C_{i-j}=\frac{(-1)^{j-1}\varrho^{-j}}{(j-1)!}$, $D_{N_\mathrm{FJ}+N_\mathrm{BS}+1-i}=\frac{(N_\mathrm{FJ}-1)(N_\mathrm{FJ}-2)....(N_\mathrm{FJ}+1-i)}{(i-1)!}(1-\varrho)^{N_\mathrm{FJ}-i}$ and $\mathrm{Li}_2(x)=\sum_{k=1}^{\infty}\frac{x^k}{k^2}$ is the Dilogarithm function {\cite {table}}.

Therefore, our closed-form expression for the asymptotic ESSR of the optimized WFJ network is given by
\begin{eqnarray}\label{ESSR_high}
\overline{R_s^{\infty}}=\frac{1}{2 \ln 2}\Big(2I_{1,1}+I_{1,2}-I_{1,3}-2 \ln 2\Big).
\end{eqnarray}
Using (\ref {power_slope}), the high SNR slope and the high SNR power offset are given by
\begin{eqnarray}\label{slope_fj}
S_{\infty}{\hspace{-3mm}}&=&{\hspace{-3mm}}1~~~~\mathrm{and}~~~~\\ \label{offset_fj}
L_{\infty}{\hspace{-3mm}}&=&{\hspace{-3mm}}-\log_2 \mu_{\mathrm{mr}} -\frac{I_{1,2}}{2 \ln 2}+\frac{I_{1,3}}{2 \ln 2}+
\frac{\mathcal{C}}{\ln 2}+1.
\end{eqnarray}
To characterize the impact of $N_\mathrm{FJ}$ and $N_\mathrm{BS}$ on the ESSR power offset in \eqref{offset_fj}, we consider  $\gamma_\mathrm{br} \gg \gamma_\mathrm{fr}$ and approximate $\ln (1-\frac{\gamma_\mathrm{fr}}{\gamma_\mathrm{br}})\approx -\frac{\gamma_\mathrm{fr}}{\gamma_\mathrm{br}}$ and $\ln (1+\frac{2\gamma_\mathrm{fr}}{\gamma_\mathrm{br}})\approx \frac{2\gamma_\mathrm{fr}}{\gamma_\mathrm{br}}$. Since $\gamma_\mathrm{fr}$ and $\gamma_\mathrm{br}$ are independent, using {\it Lemma 2.9} in {\cite {randomMat}} results in
\begin{align}
&I_{1,2} \approx -\mathbb{E}\Big\{\gamma_\mathrm{fr}\Big\}~\mathbb{E}
\Big\{\frac{1}{\gamma_\mathrm{br}}\Big\}=-\frac{N_\mathrm{FJ} \overline{\gamma}_\mathrm{fr}}{(N_\mathrm{BS}-1)\overline{\gamma}_\mathrm{br}}~~~\mathrm{and}\nonumber\\
&I_{1,3} \approx\frac{2N_\mathrm{FJ} \overline{\gamma}_\mathrm{fr}}{(N_\mathrm{BS}-1)\overline{\gamma}_\mathrm{br}}.
\end{align}
As such, the ESSR power offset is derived as
\begin{eqnarray}\label{poweroffset_OWFJ}
L_{\infty}=-\log_2 \mu_\mathrm{mr}+\frac{3 N_\mathrm{FJ}\overline{\gamma}_\mathrm{fr}}{(N_\mathrm{BS}-1)\overline{\gamma}_\mathrm{br}\ln 2}+\frac{\mathcal{C}}{\ln 2}+1.
\end{eqnarray}
Expression (\ref {poweroffset_OWFJ}) shows that the power offset of WFJ depends on  both $N_\mathrm{FJ}$ and $N_\mathrm{BS}$. As expected, we see that increasing the number of BS antennas $N_\mathrm{BS}$ decreases the ESSR power offset which corresponds to an increase in the ESSR. Interestingly, our results reveal that increasing the number of FJ antennas $N_\mathrm{FJ}$ actually increases the ESSR power offset which corresponds to a decrease in the ESSR.

\begin{figure}[t] \label{1}
  \begin{center}
    \includegraphics[width=3.30in,height=2.8in]{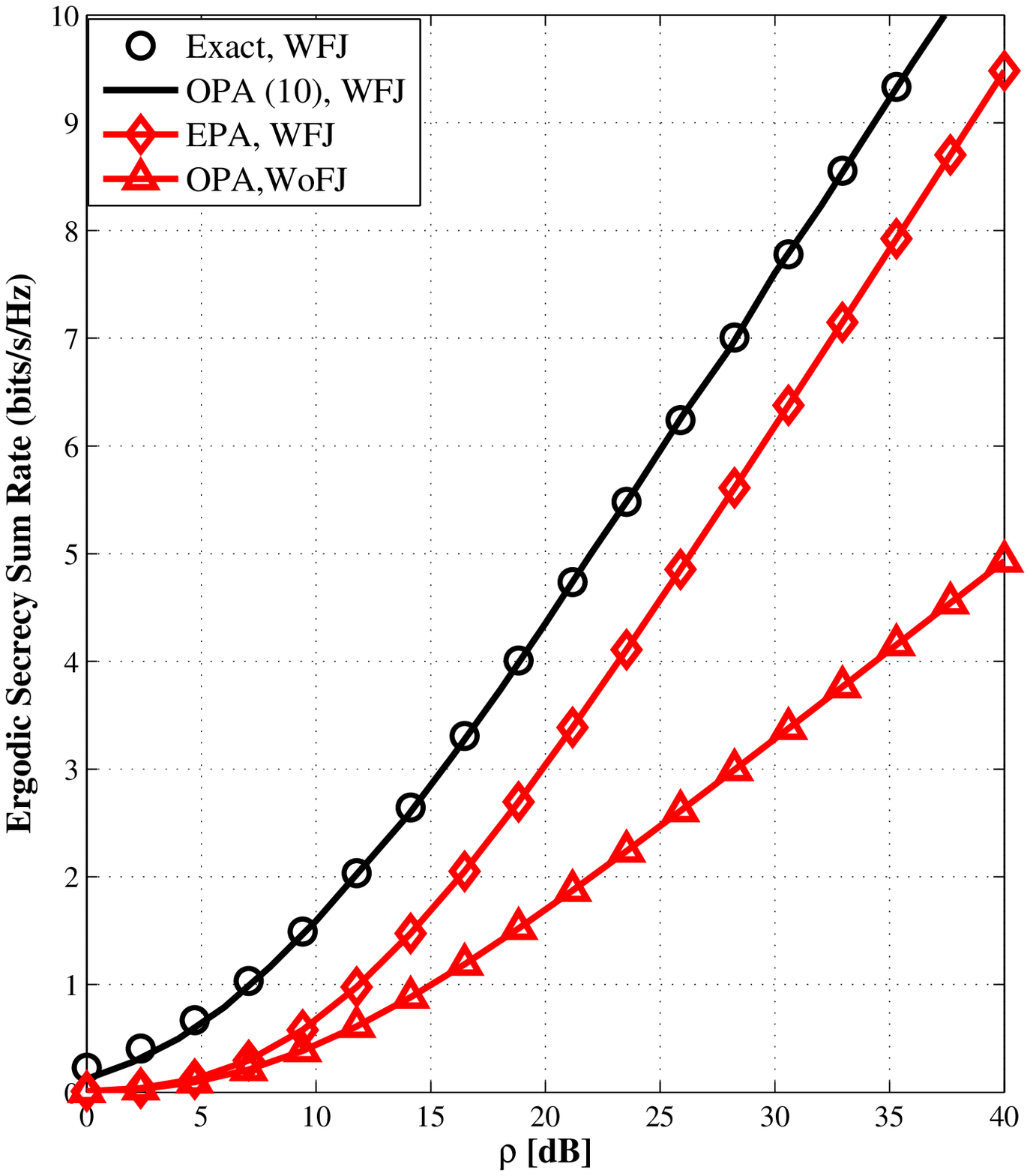} 
    \vspace{-3mm}\caption{ESSR versus transmit SNR, where $N_{\mathrm{BS}}=64$, $N_\mathrm{FJ}=1$, $\mu_{\mathrm{br}}=\mu_{\mathrm{mr}}=1$, $\mu_{\mathrm{fr}}=4$.}
  \vspace{-3mm}\label{fig1}\end{center}
\end{figure}

\begin{figure}[t] \label{1}
  \begin{center}
    \includegraphics[width=3.30in,height=2.8in]{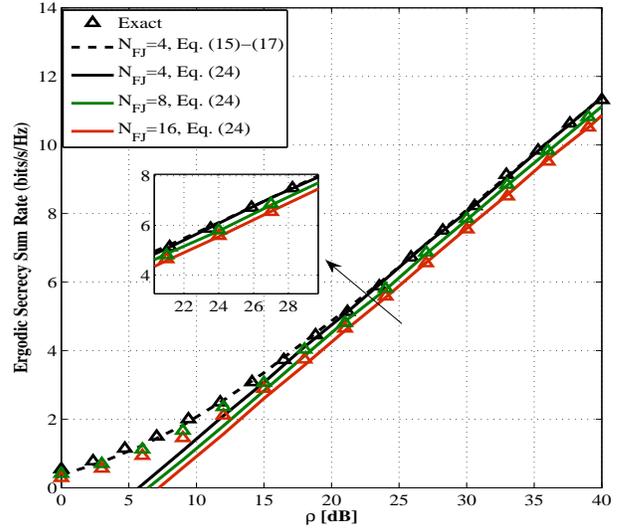} 
    \vspace{-3mm}\caption{ESSR versus transmit SNR for different number of antennas at the FJ, where $N_{\mathrm{BS}}=256$, $\mu_{\mathrm{br}}=\mu_{\mathrm{mr}}=1$, $\mu_{\mathrm{fr}}=4$.}
  \vspace{-3mm}\label{fig1}\end{center}
\end{figure}

{\vspace{-2mm}}
\section{Numerical Examples}
This section provides numerical examples to verify the accuracy of the derived ESSR expressions. We compare our LSMA-based ESSR performance with the exact ESSR where the OPA is numerically evaluated for finite numbers of BS and FJ antennas using the bisection method. We also evaluate the equal power allocation (EPA) between the BS and MU (i.e., $\lambda=0.5$) and the scenario without friendly jamming (WoFJ) where the power is distributed optimally between the users.

To verify the accuracy of the derived LSMA-based ESSR expressions in (\ref {ESSR_large})--(\ref {ESSR_large3}) and \eqref{ESSR_high}, we conduct the following simulations in Figs. 2 and 3 {\cite{zhang0}}: For simplicity and without loss of generality, we assume that the BS, the MU, the relay and the FJ are located at ($-1$, 0), (1, 0), (0, 0) and (0.3, 0.4), respectively.  Moreover, the path loss factor is $\alpha=2$.

Fig. 2 shows the ESSR versus $\rho$ in dB for $N_\mathrm{BS}=64$ and $N_\mathrm{FJ}=1$. As can be observed from the figure: 1) The ESSR using the OPA solution in (\ref {sol_J0}) is sufficiently tight at medium and high transmit SNRs, 2) The SNR gap between the optimized network and EPA is approximately 4.3 dB compared to WFJ when the ESSR is 5 bits/s/Hz, and 3) The optimized WFJ significantly improves the ESSR performance compared to the optimized WoFJ. For example, for a target ESSR of 5 bits/s/Hz, the SNR advantage of WFJ over WoFJ is approximately 18 dB which reveals the clear advantage of using a FJ to establish an energy-efficient secure network.

Fig. 3 depicts the ESSR versus $\rho$ in dB for $N_\mathrm{BS}=256$ and different values of $N_\mathrm{FJ}=4,~8,~16$. We set $T=T'=20$. As can be seen from the figure, the Monte Carlo simulation of the exact ESSR matches precisely with the ESSR expression in (\ref {ESSR_large})--(\ref {ESSR_large3}) and is well-approximated by the asymptotic ESSR expression in (\ref {ESSR_high}) in the high SNR regime. Furthermore, we note that the ESSR decreases with increasing $N_\mathrm{FJ}$ which can be explained by the fact that the high-SNR power offset  $L_{\infty}$ that was derived in (\ref {poweroffset_OWFJ}) increases with increasing $N_\mathrm{FJ}$.

\begin{figure}[t] \label{1}
  \begin{center}
    \includegraphics[width=3.30in,height=2.8in]{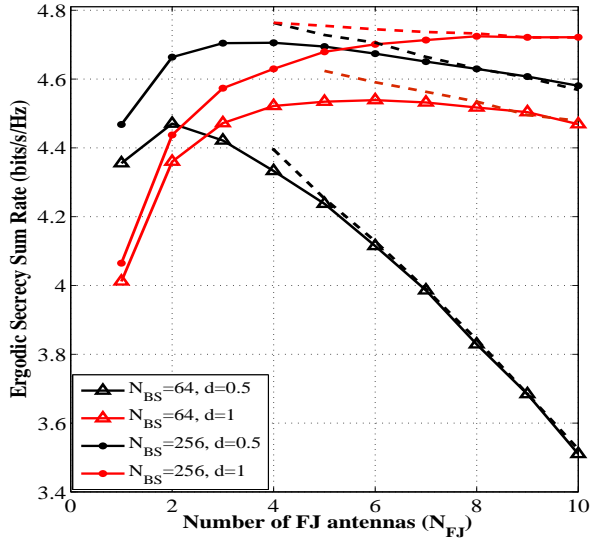} 
    \vspace{-3mm}\caption{Effect of distance between the FJ and relay on the ESSR for different number of FJ antennas $N_{\mathrm{FJ}}$, where  $\mu_{\mathrm{br}}=\mu_{\mathrm{mr}}=1$, $n=2$, $\rho=20$ dB. The dashed lines  correspond to the asymptotic SNR analysis from Eq. (\ref {ESSR_high}).}
  {\vspace{-2mm}}\label{fig1}\end{center}
  \end{figure}

In Fig. 4, we examine the effect of the FJ location and the number of FJ and BS antennas on the ESSR. The figure highlights that for a given FJ location and number of BS antennas, there is an optimal number of FJ antennas that maximizes the ESSR. For example, when $N_\mathrm{BS}=64$ and $d=0.5$, the ESSR is maximized when $N_\mathrm{FJ}=2$. For a fixed FJ location, we see that $N_\mathrm{FJ}$ must increase with increasing $N_\mathrm{BS}$ to achieve a higher ESSR. This is because the received SINR at the relay increases with  increasing $N_\mathrm{BS}$ with MRT at the BS. By employing a FJ with more antennas, the information leakage to the relay can be reduced.

{\vspace {-2mm}}
\section{Conclusion}
We examined the OPA and ESSR of a two-way relaying network including a BS, a MU, an untrusted relay and a FJ. We considered the BS and FJ are equipped with multiple antennas denoted as $N_\mathrm{BS}$ and $N_\mathrm{FJ}$, respectively, and derived the OPA between the BS and the MU such that the instantaneous secrecy sum rate of the network is maximized. Based on the OPA solution, new closed-form ESSR expressions were derived for the optimized WFJ network in the LSMA regime. Our asymptotic ESSR expressions highlighted that the ESSR of WFJ depends on the number of antennas $N_\mathrm{BS}$ and $N_\mathrm{FJ}$. Numerical results revealed that exploiting a FJ improves the ESSR significantly compared to without friendly jamming.

\appendices

\vspace{-1mm}
\section{}\label{appC}\vspace{-1mm}
To evaluate $I_1$, we first derive the cumulative distribution function (cdf) of $\gamma_\mathrm{fb}{\stackrel{\tiny \Delta}{=}}\frac{\gamma_\mathrm{fr}}{\gamma_\mathrm{br}}$ as follows
\begin{align}\label{chi_sqyared}
&f_{\gamma_{\mathrm{fb}}}(x)=\frac{\partial }{\partial x} \int_0^{\infty}\int_0^{\beta x} f_{\gamma_\mathrm{fr}}(\alpha)
f_{\gamma_\mathrm{br}}(\beta)~\mathrm{d} \alpha \mathrm{d} \beta\nonumber\\
&=\frac{\Gamma(N_\mathrm{FJ}+N_\mathrm{BS})}{\Gamma(N_\mathrm{FJ})\Gamma(N_\mathrm{BS})}\times \Big(\frac{\overline{\gamma}_\mathrm{fr}}{\overline{\gamma}_\mathrm{br}}\Big)^{N_\mathrm{BS}}{\hspace{-4mm}}\times {\hspace{-1mm}} \frac{x^{N_\mathrm{FJ}-1}}{(x+\frac{\overline{\gamma}_\mathrm{fr}}{\overline{\gamma}_\mathrm{br}})^{N_\mathrm{FJ}+N_\mathrm{BS}}},
\end{align}
where the last expression follows from Leibniz integral rule and substituting the probability density function (pdf) of $\gamma_\mathrm{fr}$ and $\gamma_\mathrm{br}$, and using {\cite[Eq. (3.381.4)] {table}}. Due to the fact that $\gamma_\mathrm{mr}>1$ and $\gamma_\mathrm{fb}<1$, the cdf of $\gamma_\mathrm{BS}$ in (\ref {gammas}) is obtained as
\begin{eqnarray}\label{f_omega0}
F_{\gamma_{\mathrm{BS}}}(\gamma){\hspace{-3mm}}&=&{\hspace{-3mm}}\Pr\Big\{\gamma_{\mathrm{BS}}<\gamma\Big\}=
\mathbb{E}_{\gamma_\mathrm{fb}}\Big\{F_{\gamma_\mathrm{mr}}\Big(\frac{\gamma (1+2 \gamma_\mathrm{fb})}{1-\gamma_\mathrm{fb}}\Big)\Big\}\nonumber\\
&{\stackrel{\tiny (a)}{=}}&{\hspace{-3mm}}1-\mathbb{E}_{\gamma_\mathrm{fb}}\Big\{e^{-\frac{\gamma}{\overline{\gamma}_\mathrm{mr}}
\Big(\frac{1+2\gamma_\mathrm{fb}}{1-\gamma_\mathrm{fb}}\Big)}\Big\}\nonumber\\
&{\stackrel{\tiny (b)}{=}}&{\hspace{-3mm}} 1- e^{\frac{2\gamma}{\overline{\gamma}_\mathrm{mr}}}\frac{\Gamma(N_\mathrm{FJ}+N_\mathrm{BS})}{\Gamma(N_\mathrm{FJ})\Gamma(N_\mathrm{BS})}\times \Big(\frac{\overline{\gamma}_\mathrm{fr}}{\overline{\gamma}_\mathrm{br}}\Big)^{N_\mathrm{BS}}\nonumber\\
&&{\hspace{-3mm}}\int_0^1e^{-\frac{3\gamma}{\overline{\gamma}_\mathrm{mr}}\times \frac{1}{1-z}}
\frac{z^{N_\mathrm{FJ}-1}}{(z+\frac{\overline{\gamma}_\mathrm{fr}}{\overline{\gamma}_\mathrm{br}})^{N_\mathrm{FJ}+N_\mathrm{BS}}}~\mathrm{d}z,
\end{eqnarray}
where $(a)$ and $(b)$ follow from substituting the cdf of $\gamma_\mathrm{mr}$ and the pdf of $\gamma_\mathrm{fb}$, respectively.
By using a change of variables $u=\frac{1}{1-z}$ and then applying the partial fraction decomposition from {\cite[Eq. (2.102)] {table}}, the cdf of $\gamma_\mathrm{BS}$ is formed. By using the obtained cdf and {\cite[Eq. (3.352.4)]{table}}, $I_1$ is expressed as
\begin{eqnarray}
{\hspace{-3mm}}&\mathbb{E}&\hspace{-3mm}\Big\{\ln \Big(1+\gamma_\mathrm{BS}\Big)\Big\}=-\frac{\Gamma(N_\mathrm{FJ}+N_\mathrm{BS})}{\Gamma(N_\mathrm{FJ})\Gamma(N_\mathrm{BS})} \Big(\frac{\overline{\gamma}_\mathrm{fr}}{\overline{\gamma}_\mathrm{br}}\Big)^{N_\mathrm{BS}} \nonumber\\
{\hspace{-3mm}}&\times&{\hspace{-5mm}}\sum_{i=1}^{N_\mathrm{FJ}+N_\mathrm{BS}}{\hspace{-3mm}}A_i\int_1^{\infty}\frac{e^{\frac{3u-2}{\overline{\gamma}_\mathrm{mr}}}\mathrm{Ei}
(-\frac{3u-2}{\overline{\gamma}_\mathrm{mr}})}{(u-\frac{1}{\varrho})^i}~\mathrm{d}u.
\end{eqnarray}
By using a change of variables $v=\frac{3u-2}{\overline{\gamma}_\mathrm{mr}}$, the result in (\ref {Ei_x}) along with [11, Eq. (3.352.2)] and [11, Eq. (3.351.4)], the result in (\ref {ESSR_large}) can be achieved after simple manipulations.
{\vspace{-3mm}}
\section{}\label{appD}
Using [11, Eq. (4.352.2)], $I_{1,1}$ is obtained as (\ref {I_11}). Furthermore, using the pdf of $\gamma_\mathrm{fb}$ in (\ref {chi_sqyared}), $I_{1,2}$ is derived as
\begin{eqnarray}\label{diff1}
{\hspace{-3mm}}\mathbb{E}\Big\{\ln(1-{\gamma_\mathrm{fb}})\Big\}
{\hspace{-3mm}}&=&{\hspace{-3mm}}\frac{\Gamma(N_\mathrm{FJ}+N_\mathrm{BS})}{\Gamma(N_\mathrm{FJ})\Gamma(N_\mathrm{BS})}
\Big(\frac{\overline{\gamma}_\mathrm{fr}}{\overline{\gamma}_\mathrm{br}}\Big)^{N_\mathrm{BS}}\nonumber\\
&\times&{\hspace{-3mm}}\frac{1}{(-1)^{N_\mathrm{BS}+1}}{\hspace{-3mm}}\sum_{i=1}^{N_\mathrm{FJ}+N_\mathrm{BS}}{\hspace{-3mm}}B_i{\hspace{-1mm}}\int_{0}^1\frac{\ln (u)}{(u-\varrho)^i}~\mathrm{d}u.
\end{eqnarray}
Using {\cite[Eq. (2.727.1)]{table}} and the partial fraction decomposition of {\cite[Eq. (2.101)]{table}}, $I_{1,2}$ is computed as shown in \eqref{I12}. The third part $I_{1,3}$ can be calculated in the same way as $I_{1,2}$  and the final expression can be found in \eqref{I13}.\\

{\vspace{-0mm}}
~~~~~~~~~~~~~~~~~~~~~{\bf Acknowledgements}

The authors deeply thank Prof. Lajos Hanzo for useful discussions and constructive comments to improve the paper.

\end{document}